# From OA colors to observable access states: A commentary on the 2026 OpenAlex proposal

Abdelghani Maddi
abdelghani.maddi@cnrs.fr
*Sorbonne Université, CNRS, Groupe d'Étude des Méthodes de l'Analyse Sociologique de la Sorbonne, GEMASS, Paris, France*

## Abstract

The OpenAlex Colors Working Group proposes replacing the familiar Gold, Green, Hybrid, Bronze and Closed taxonomy with a work-level framework based on three dimensions: access, location and license. The proposal addresses genuine weaknesses in the legacy color system, particularly its dependence on journal-level characteristics, its difficulty in describing non-article outputs, and its conflation of access conditions with venue business models. This commentary supports that general direction but argues that the proposed labels should not become the primary data model. The central technical issue is that access, location and license are not best understood as three independent attributes of a work: scholarly works commonly have several versions or copies at different locations, and access status, license and version may differ from one location to another. The analysis therefore recommends treating the location/version observation as the atomic unit of OA metadata and deriving work-level labels, legacy colors and user-facing summaries from those observations. Particular attention is given to license semantics, version identification, the normative assumptions embedded in a "best-available" hierarchy, the coarse Publisher/Other Platform distinction, temporal change, provenance, compound works, and backward compatibility. Existing standards and infrastructure (including NISO/ALPSP Journal Article Versions, COAR controlled vocabularies, FAIR provenance principles, and OpenAlex's own current location objects) already provide many of the components needed for such an approach. The result would preserve the simplicity sought by the Working Group while making the underlying data model more extensible, auditable and suitable for funder compliance, bibliometric research and future changes in scholarly communication.

**Keywords:** Open access; OpenAlex; Unpaywall; scholarly communication; metadata; licensing; repositories; bibliometrics; research infrastructures

## 1. Introduction

Open-access (OA) colors have become a compact vocabulary for describing how scholarly works can be accessed. The taxonomy popularized by Unpaywall and the 2018 "State of OA" study distinguishes Gold, Green, Hybrid, Bronze and Closed access, with the categories subsequently reused in bibliometric databases, institutional monitoring and research-policy workflows (Piwowar et al., 2018). Its success is also the source of its present difficulty: a pragmatic classification developed primarily around journal articles has become infrastructure for a much wider range of objects, policies and analytical questions.

The OpenAlex Colors Working Group responds to this problem by proposing a new work-level classification based on Access (Free-to-read or Closed), Location (Publisher or Other Platform), and License (Open, Closed/Restrictive, or Unknown). These dimensions generate labels such as Publisher Open, Platform Open, Publisher Public, Platform Public, Publisher Reserved, Platform Reserved and Closed. The report explicitly separates access status from venue business models and recommends preserving granular metadata and a transition path from legacy colors (Alperin et al., 2026, pp. 7-12).

This is a strong conceptual correction. The move away from journal-centric categories, the separation of observable access properties from economic models, and the insistence on work-level metadata all respond to real limitations of the legacy system. The main argument developed here is narrower: the proposal should go one step further and distinguish the underlying data model from the labels derived from it. A robust OA infrastructure should store observations about particular versions at particular locations and calculate simplified work-level labels as views. If the labels themselves become the core representation, several of the ambiguities of the color system risk being recreated under new names.

# 2. Scope and analytical approach

The primary object of analysis is the July 2026 consultation draft of the Working Group report. A public version is now available on Zenodo (Alperin et al., 2026). Because OpenAlex is an actively changing infrastructure, the analysis also considers OpenAlex documentation available on 18 September 2026, together with relevant standards and policy frameworks: NISO/ALPSP terminology for journal-article versions, COAR controlled vocabularies for access rights and version types, cOAlition S requirements for funder compliance, Creative Commons licensing documentation, FAIR principles for provenance, and established literature on OA classification and self-archiving.

This is a conceptual and technical commentary rather than an empirical validation study. It does not estimate classification error rates or compare a sample of OpenAlex records against publisher and repository websites. Its purpose is instead to test whether the proposed model is internally coherent, interoperable with existing standards, sufficiently expressive for the use cases identified by the Working Group, and resilient to foreseeable changes in scholarly communication.

**Table 1. Main contributions of the proposal and issues that still require specification**

| Proposal element | Contribution | Remaining design question |
|---|---|---|
| **Work-level classification** | Reduces dependence on current venue policy and accommodates non-journal outputs. | The same work can have several versions and locations with different access and licensing states. |
| **Access / Location / License** | Makes important properties explicit instead of hiding them in a color label. | These properties are attached naturally to a particular copy/location, not always to the work as a whole. |
| **All-Versions Rule** | Recognizes that multiple access routes can coexist. | A normalized location-level model is cleaner than concatenated work-level labels. |
| **Best-Available label** | Provides a convenient single summary for interfaces and simple analyses. | The ranking embeds a preference order and should be treated as configurable display logic, not the core data model. |
| **Business models separated** | Avoids treating Gold/Hybrid as both access and economic categories. | A parallel business-model layer must be developed without creating a long functional gap. |
| **Legacy mapping** | Acknowledges downstream dependencies and the cost of migration. | The mapping is not one-to-one and therefore requires explicit rules, versioning and regression tests. |

# 3. What the proposal gets right

## 3.1 Work-level classification is the correct default

The report is persuasive in arguing that a publication should not inherit its OA status primarily from the current policy of the journal in which it appeared. Journal policies change, journals flip between access models, and historical backfiles may have very different access properties from current issues. A work-level approach therefore better describes the object a user is actually trying to read or reuse (Alperin et al., 2026, pp. 4-5). This direction is also consistent with the broader move toward evaluating and describing research outputs on their own terms rather than using venue-level proxies. DORA, although concerned with research assessment rather than OA classification, makes the analogous methodological point that journal-level attributes should not be used as substitutes for the properties of individual outputs (DORA, 2012).

The work-level turn is especially important for preprints, dissertations, datasets, software and books. The legacy Gold/Green distinction assumes a publisher-repository relationship that often does not exist for these objects. An output-agnostic framework is therefore necessary if OpenAlex is to remain a general scholarly knowledge graph rather than an article index with extensions (Priem, Piwowar, & Orr, 2022).

## 3.2 Access status and business model should be distinct metadata layers

The report is also right to separate the question “How can this work be accessed and reused?” from “How was its publication financed or organized?” Gold has accumulated several meanings in practice: publisher-hosted OA, fully OA journals, APC-funded publishing and, in some contexts, simply the preferred route to openness. These are not equivalent properties. Diamond journals illustrate the problem particularly well: they are free to readers and authors, but their governance and funding arrangements vary and cannot be inferred from the access state of a single article (Becerril et al., 2021). Conversely, fee-based OA can occur in fully OA or hybrid settings, and cost information requires dedicated sources such as OpenAPC rather than inference from access alone (OpenAPC, 2026).

A complementary business-model taxonomy is therefore preferable to embedding funding assumptions in access labels. The important implementation condition is sequencing: the new access taxonomy and the parallel economic metadata layer should be planned together so that users who currently rely on Gold, Hybrid or Diamond for budgeting and negotiation do not lose information during transition.

## 3.3 Granular metadata and backward compatibility are essential

The Working Group explicitly states that power users must continue to access the underlying variables rather than being forced to rely on a simplified label, and it recommends a transition period in which legacy colors remain queryable (Alperin et al., 2026, pp. 6-7, 11-12). These are not secondary implementation details. A classification used by funders, libraries, bibliometricians and commercial databases becomes an interoperability contract. Any redesign should therefore preserve machine-readable evidence, publish conversion rules, and allow downstream systems to migrate at different speeds.

# 4. The central modeling issue: the three dimensions are not fully independent at work level

The report presents Access, Location and License as three independent dimensions of a work. In practice, however, access and license are properties of a particular copy at a particular location. A single scholarly work may be closed at the publisher, openly licensed as an accepted manuscript in an institutional repository, and available as a differently licensed preprint on a preprint server. The work does not have one location and one license; it has a set of location-specific observations.

The report itself moves toward this conclusion when discussing the All-Versions Rule. It recommends storing classifications for each location and allowing combined descriptions such as “Publisher Public and Platform Open” (Alperin et al., 2026, pp. 8-9). Once classification is location-specific, however, the basic unit of the model is no longer a single three-dimensional work record. It is a one-to-many structure in which a work has multiple location/version records, each with its own access and licensing attributes.

OpenAlex's current public documentation already reflects much of this normalized structure. A location is documented as a place where a version of a work is available, and each location can carry its own OA flag, license, version and source. Works aggregate these locations and expose pointers such as primary_location and best_oa_location (OpenAlex, 2026a, 2026b). This is a significant design clue: the proposed reform may require less invention of a new core schema than clarification of the semantic layer derived from an already granular location model.

A more robust formulation would therefore treat the individual locations of a work as the primary observations. For each location, the data model should preserve the version available, the hosting source, whether the copy is free to read, its exact license, and relevant dates. Simplified categories such as “Publisher Open” or “Platform Public” can then be derived from these underlying records. In this approach, the classification summarizes the underlying evidence rather than replacing it.

## 5. License semantics require greater precision

The proposed Open / Closed / Unknown license dimension is useful for simple reporting but becomes problematic if treated as primary metadata. The consultation draft notes that OpenAlex has treated all Creative Commons licenses as open, yet the new "Reserved" categories use CC BY-NC-ND as an example of a restrictive license (Alperin et al., 2026, pp. 7-8). This exposes a genuine definitional choice rather than a minor inconsistency.

Different communities operationalize "open" differently. The Budapest Open Access Initiative links OA to broad rights to read, copy, distribute and reuse. The Open Definition treats CC BY and CC BY-SA as conformant while excluding NonCommercial and NoDerivatives licenses from its definition of open content. Creative Commons, meanwhile, accurately describes CC BY-NC-ND as permitting redistribution but prohibiting commercial use and adaptations. cOAlition S adds a policy layer: CC BY is the default requirement, with CC BY-SA and CC0 accepted and CC BY-ND permitted only by exception in its implementation guidance (Budapest Open Access Initiative, 2002; Creative Commons, 2013, 2026; cOAlition S, 2019).

OpenAlex's current location model already preserves normalized license values and canonical license identifiers when available (OpenAlex, 2026b). The revised classification should retain this granular layer and treat "open", "restricted" or "policy-compliant" as derived predicates whose rule set is explicit and versioned. A user interested in legal reuse, a funder checking compliance and a bibliometrician measuring free-to-read availability do not necessarily need the same binary classification. Preserving the exact license prevents a system-wide semantic decision from erasing distinctions that are already available in the source data.

## 6. Version should be a first-class variable

Version is absent from the three proposed headline dimensions even though several of the report's own use cases depend on it. Funder compliance may turn on whether the accessible copy is a submitted manuscript, an Author Accepted Manuscript (AAM), or the Version of Record (VoR). The report explicitly acknowledges this requirement when discussing funders and multiple versions (Alperin et al., 2026, pp. 6, 8).

There is no need to invent a new vocabulary. NISO/ALPSP Journal Article Versions defines a sequence including Author's Original, Submitted Manuscript Under Review, Accepted Manuscript, Proof, Version of Record, Corrected Version of Record and Enhanced Version of Record (NISO, 2008). COAR reuses this terminology in a multilingual controlled vocabulary for repositories (COAR, 2022b). cOAlition S likewise distinguishes AAM and VoR in its compliance routes and requires repositories to expose the deposited version, its Open Access status and its license in metadata (cOAlition S, 2020).

OpenAlex already records a version field on each location, using publishedVersion, acceptedVersion or submittedVersion where it can determine the version (OpenAlex, 2026b). The revised classification should keep version as a first-class location-level variable rather than infer it from an access label. For other output types, the model should permit a broader or extensible version vocabulary. This is preferable to forcing all scholarly objects into article-specific terminology while still recognizing that version is indispensable wherever version-specific policies exist.

## 7. "Best available" is useful interface logic, not a measure of openness

To provide one simple label per work, the Working Group proposes a hierarchy in which Publisher Open is preferred to Platform Open, followed by Publisher Public, Platform Public, Publisher Reserved, Platform Reserved and Closed. The report correctly notes that this introduces a normative judgment (Alperin et al., 2026, p. 8). The problem is not that a user interface needs a default; it is that the default can be mistaken for a conceptual ranking of OA routes.

One possible rationale for ranking “Publisher Open” above “Platform Open” is that a publisher-hosted copy will often correspond to the Version of Record, whereas a repository copy may be a preprint or an accepted manuscript. If this is the intended rationale, however, the relevant distinction is the version of the work rather than its hosting location. The proposed framework defines Location independently from Version, and the report itself acknowledges that scholarly works may exist as preprints, accepted manuscripts, and Versions of Record across multiple locations. Publisher websites may also host preprint versions. A hierarchy that systematically places Publisher Open above Platform Open therefore risks using location as a proxy for version status. Recording version explicitly would make it possible to distinguish, for example, an openly accessible Version of Record from an openly accessible accepted manuscript without assuming that one hosting environment is intrinsically more open than another.

Which copy should be preferred may then depend on the use case: a reader seeking the definitive citable text may prefer the Version of Record, whereas a funder monitoring repository-deposit compliance may specifically need the deposited manuscript. These are differences in version and use case, rather than intrinsic differences in openness.

OpenAlex already exposes a best_oa_location and, as documented in September 2026, ranks publisher-hosted locations ahead of repository-hosted locations before considering version and other tie-breakers (OpenAlex, 2026a, 2026b). That ranking can remain a practical display policy, but it should be named and documented as such. A field such as preferred_display_location or default_oa_location would be less semantically loaded than treating the resulting label as the work's definitive OA category. Ideally, advanced users should be able to apply alternative ranking functions.

# 8. “Publisher” versus “Other Platform” is too coarse for a durable location vocabulary

The Working Group openly recognizes that “publisher” becomes ambiguous for datasets, theses, preprints and other non-article outputs, and that “Other Platform” groups together heterogeneous services (Alperin et al., 2026, pp. 9-10). This is not merely a terminology problem because one of the three headline dimensions depends on the distinction.

A more extensible approach is to store the source or provider and a controlled provider type at the location level, for example publisher platform, institutional repository, disciplinary repository, preprint server, data repository, general-purpose repository, personal or project website, and other. These detailed types can then be collapsed into Publisher / Platform if a simplified interface requires it. The principle is the same as for licenses: retain the most specific interoperable observation and derive broader categories rather than discarding information at ingestion.

This also avoids making the publisher/repository dichotomy carry more semantic weight than it can bear. Crossref relation metadata already distinguishes relations such as preprint, manuscript, manifestation and version, while COAR maintains vocabularies for repository-oriented resource and version description (Crossref, 2020; COAR, 2022b). OpenAlex can align with these existing semantics rather than building a new binary boundary whose difficult cases are known in advance.

# 9. Time, provenance and uncertainty should be designed in from the start

## 9.1 Access is time-dependent, not a timeless attribute

The report identifies the absence of a temporal dimension as a weakness of the current system: works may become open after an embargo, disappear from free access, or move through several states. Yet full temporal tracking is deferred because it would complicate implementation (Alperin et al., 2026, pp. 5, 10-11). The issue is substantive rather than marginal: publisher self-archiving policies have long varied by version, location and delay, making "when" an important part of Green OA (Laakso, 2014).

OpenAlex's current infrastructure already provides part of the required foundation. As documented on 18 September 2026, location records expose provenance and an ingested_at timestamp, while the underlying OA data retains an oa_date indicating when a copy first became available at a location; oa_date is currently exposed through the Unpaywall-compatible view rather than the OpenAlex work object (OpenAlex, 2026a, 2026b). These fields do not constitute a complete access history, but they show that temporal and provenance information can be attached to location-level records.

A first implementation of the revised classification therefore need not reconstruct the complete access history of every work. The important requirement is to preserve and, where useful, extend location-level dates so that users can distinguish when a copy was ingested or observed from when it is known to have become openly available. A last-verified date could also be retained where repeated verification is performed. This would support reproducible longitudinal analysis while keeping the model compatible with more complete historical tracking in the future.

### 9.2 Provenance and confidence are equally important

A descriptive classification based on observed properties also needs to preserve how those properties were obtained. A license detected in machine-readable metadata, a license inferred from publisher text, a repository record harvested by OAI-PMH and an HTTP accessibility check do not have the same evidential status. The FAIR principles explicitly identify detailed provenance and community standards as conditions of reusable metadata, and the W3C PROV family provides a general framework for recording the origin and processing history of data (Wilkinson et al., 2016; W3C, 2013).

OpenAlex already exposes provenance on location records, which provides a useful foundation (OpenAlex, 2026b). The revised model should preserve that provenance and make it sufficiently specific for access and license assertions; where inference rather than direct metadata is involved, an evidence or confidence code would also be useful. This would make disagreements diagnosable and allow downstream users to choose their own tolerance for uncertainty. "Unknown license" in particular should distinguish "no license asserted at the observed location" from "a license exists but could not be parsed or mapped."

## 10. Compound works need a mixed state, not forced closure

For works with components, the report proposes applying the framework at component level while describing the container as Closed unless all components share the same open classification (Alperin et al., 2026, p. 9). This rule is easy to implement but discards substantial information. A book with nineteen openly accessible chapters and one closed chapter would become indistinguishable at container level from a book with no open chapters.

A compound object should instead be able to carry a Mixed or Partial summary accompanied by counts or proportions of components by state. The precise aggregation rule can vary by object type, but the general principle should be monotonic: adding open components should not leave the container's summary indistinguishable from an entirely closed object. The same issue arises for edited volumes, proceedings, datasets with restricted files, and compound digital objects.

## 11. Business-model metadata should be parallel, but not postponed indefinitely

Conceptual separation does not imply operational separation. Libraries and funders often use OA categories precisely because they need to understand expenditure, agreements and non-APC models. The Working Group appropriately recommends a separate initiative for APC-funded, Diamond,

Subscribe-to-Open and transformative arrangements (Alperin et al., 2026, pp. 10, 12). Evidence from the Diamond OA literature and OpenAPC also demonstrates why these questions require sources beyond the access state of a work (Becerril et al., 2021; OpenAPC, 2026).

The transition plan should therefore include a minimum viable business-model layer or a clearly synchronized roadmap. Otherwise, removing venue-level semantics from the OA label may improve conceptual purity while reducing immediate utility for some institutional workflows. A modular architecture can avoid this trade-off: access observations, venue business-model metadata and cost/agreement data can remain separate entities while being linked for analysis.

## 12. Backward compatibility needs explicit, versioned transformation rules

The appendix mapping between new labels and legacy colors is useful but not a reversible crosswalk. Publisher Open corresponds to either Gold or Hybrid depending on venue context; Platform Open and Platform Public split the former Green category; Reserved categories have no legacy equivalent; and a work with multiple locations can satisfy more than one proposed label simultaneously. A downstream system therefore cannot reconstruct the old status from the new label without additional rules and venue information.

This is particularly important because OpenAlex itself is evolving during the consultation. The July draft notes that Gold includes Diamond under the then-current approach, while the OpenAlex help documentation accessed on 18 September 2026 lists Diamond as a distinct oa_status alongside Gold, Green, Hybrid, Bronze and Closed (Alperin et al., 2026, p. 3; OpenAlex, 2026a). The difference does not invalidate the report; it demonstrates why taxonomy versions and observation dates must be explicit.

A production migration should therefore publish: (1) a versioned transformation specification, not only a prose table; (2) test fixtures covering ambiguous and multi-location cases; (3) a schema-version field; (4) a defined period of dual exposure for old and new fields; and (5) release notes that document changes to the derivation rules. Legacy colors can then be maintained as compatibility views even after they cease to be the preferred conceptual vocabulary.

## 13. A location/version observation model

The critique above can be summarized in one architectural recommendation: make the most granular observable record the primary data layer, and make simplified taxonomies derived views. For OA metadata, the natural observation unit is a version of a work at a specific location, with relevant temporal information where available. Table 2 gives a minimal illustrative schema. It is intentionally aligned with concepts already present in OpenAlex locations and with external vocabularies rather than proposing an entirely new data model.

**Table 2. Suggested minimum fields for an atomic OA location/version observation**

| Field | Purpose | Illustrative values | Status |
|---|---|---|---|
| **work_id** | Links the observation to the scholarly work. | OpenAlex work ID / DOI | Core |
| **location / source** | Identifies where the copy is hosted. | URL, source ID, repository or publisher | Core |
| **provider_type** | Preserves a detailed hosting category. | publisher; institutional repository; preprint server; data repository; other | Core or controlled vocabulary |
| **version** | Identifies the version available at that location. | submitted; accepted; published; corrected | Core where applicable; align with NISO/COAR |
| **is_oa / free_to_read** | Records whether full text can be read without payment or login. | true / false / unknown | Core observation |
| **license / license_id** | Stores the detected legal instrument without premature simplification. | cc-by; cc-by-nc-nd; publisher-specific; null | Already represented in the current location model |
| **access_constraints** | Captures barriers that a binary flag may hide. | registration; time-limited; token-limited; none | Recommended |
| **ingested_at / oa_date / last_verified_at** | Distinguishes ingestion, known OA availability and subsequent verification. | ISO 8601 timestamps | Partly existing; extend where useful |
| **provenance / evidence** | Records how the assertion was obtained. | embedded metadata; repository feed; page detection; manual curation | Existing field; preserve and extend evidence detail as needed |
| **confidence / mapping status** | Distinguishes direct assertions from inference or unresolved parsing. | asserted; inferred; ambiguous; unmapped | Recommended |

From this layer, OpenAlex can calculate multiple views without duplicating or losing evidence. A default human-facing status might still be Publisher Open or Platform Public. A funder-specific view could test the exact license, version and embargo conditions. A legacy view could reproduce Gold/Green/Hybrid/Bronze using a documented historical algorithm. Venue-level summaries could aggregate observations over a specified publication interval. Business-model attributes could be joined from a separate venue or agreement layer. This architecture recognizes that classification is a function over observations, not the observations themselves.

The approach is also consistent with FAIR practice: retain rich attributes, explicit licenses, provenance and community-standard vocabularies so that downstream users can recombine the data for purposes not anticipated by the original classification designer (Wilkinson et al., 2016). It would make future terminological changes significantly cheaper because changing a view would not require reinterpreting or discarding the underlying evidence.

## 14. Implementation and evaluation priorities

The Working Group recommends a phased pilot and success metrics, including downstream uptake, ambiguity reports, manual edge cases and stakeholder satisfaction (Alperin et al., 2026, p. 12). Those recommendations could be made more diagnostic by evaluating the underlying observation model as well as the labels.

- Stratify pilot records by output type: journal articles, preprints, books/chapters, theses, datasets and software.
- Oversample difficult cases: multiple OA locations, journal flips, embargoed deposits, temporary publisher access, missing or conflicting licenses, and works with several versions.
- Measure not only label agreement but evidence completeness: percentage of OA locations with a version, exact license, timestamp and provenance source.
- Test longitudinal reproducibility by re-running classifications after access changes and verifying that prior states remain interpretable.
- Publish machine-readable fixtures and transformation rules so downstream providers can validate their own implementations.
- Evaluate simplified labels separately from the raw schema: user satisfaction with a display label should not be used as evidence that the underlying metadata model is sufficiently expressive.

This would also enable a productive division between stable infrastructure and revisable policy. The location-level record can change slowly and remain backward compatible, while naming conventions, display hierarchies and policy-compliance functions can evolve more rapidly in response to community feedback.

## 15. Conclusion

The OpenAlex Colors Working Group has identified the right problem and proposes several important corrections. The legacy color system is increasingly difficult to apply consistently across diverse research outputs, journal histories and licensing arrangements. Moving to observable work-level properties, separating access from business models, preserving granular metadata and planning a backward-compatible transition are all well-founded directions.

The main point raised in this commentary concerns the role of the proposed labels themselves. Categories such as Publisher Open, Platform Open or Platform Public can provide useful summaries for users, but they should not replace the more detailed information from which they are derived. A single scholarly work may have several versions available in different places: for example, a Version of Record on a publisher website, an accepted manuscript in an institutional repository, and a preprint on a preprint server. These copies may differ in their licenses and access conditions, and their availability may also change over time. For this reason, the underlying data model should preserve information about each observed copy of a work: its location, version, access status, exact license, and, where possible, the date on which that information was observed or verified. The proposed OpenAlex categories could then be calculated from these more detailed records and presented as simplified labels when needed.

This approach would be particularly appropriate for OpenAlex because its data are reused for many different purposes. Funders, libraries, researchers and bibliometric databases do not necessarily apply the same definitions of open access or require the same versions of a work. Preserving the underlying evidence would allow these users to construct classifications suited to their own requirements while still benefiting from simple default categories provided by OpenAlex.

The proposed reform should therefore be understood not only as an opportunity to replace one classification system with another, but also as an opportunity to clarify the relationship between the underlying metadata and the categories derived from them. A durable system would preserve detailed observations first and derive classifications from them second. This would make future revisions of the taxonomy easier while reducing the risk of losing information that may become important for new policies, research questions or forms of scholarly communication.